\documentclass[aps,prb,twocolumn,showpacs,amsmath,amssymb,groupedaddress]{revtex4}
\usepackage{graphicx}
\usepackage{amssymb}
\usepackage{epstopdf}
\usepackage{dcolumn}
\usepackage{bm}
\usepackage{subfigure}
\usepackage[usenames]{color}

\begin{document}
\title{Magnetocrystalline anisotropy and uniaxiality of MnAs/GaAs(100) films}

\author{J. Magnus Wikberg$^{1}$, Ronny Knut$^{2}$, Sumanta Bhandary$^{2}$, Igor di Marco$^{2}$, Mikael Ottosson$^{3}$, Janusz Sadowski$^{4}$, Biplab Sanyal$^{2}$, P\aa l Palmgren$^{2}$, Cheuk W. Tai$^{5}$, Olle Eriksson$^{2}$, Olof Karis$^{2}$ and Peter Svedlindh$^{1}$}
\affiliation{$^{1}$Department of Engineering Sciences, Uppsala University, P.O. Box 534, SE-751 21 Uppsala, Sweden}
\affiliation{$^{2}$Department of Physics and Astronomy, Uppsala University, Box 516, 751 20 Uppsala, Sweden}
\affiliation{$^{3}$Department of Materials Chemistry, Uppsala University, P.O. Box 538, SE-751 21 Uppsala, Sweden}
\affiliation{$^{4}$MAX-lab, Lund University, P.O. Box 118, SE-221 00 Lund, Sweden}
\affiliation{$^{5}$Department of Materials and Environmental Chemistry, Arrhenius Laboratory, Stockholm University,SE-106 91 Stockholm, Sweden}

\begin{abstract}
We present an investigation of the magnetic behavior of epitaxial MnAs films grown on GaAs(100). We address the dependence of the magnetic moment, ferromagnetic transition temperature ($T_c$) and magnetocrystalline anisotropy constants on epitaxial conditions. From thorough structural and magnetic investigations, our findings indicate a more complex relationship between strain and magnetic properties in MnAs films than a simple stretch/compression of the unit cell axes. While a small increase is seen in the anisotropy constants the enhancement of the magnetic moment at saturation is significant. X-ray magnetic circular dichroism results show a behavior of the spin- and orbital-moment which is consistent with a structural transition at $T_c$. In particular, we find that the ratio of the orbital to spin moment shows a marked increase in the coexistence region of the ferromagnetic $\alpha$- and paramagnetic $\beta$-phases, a result that is well in accord with the observed increase of the $c/a$-ratio in the same temperature region. The \textit{ab initio} density functional calculations reveal that the magnetic properties are more sensitive towards change in in-plane axis as compared to a change of the out-of-plane axis, which is explained by the analysis of band structures. The effects of electron correlation in MnAs using \textit{ab initio} dynamical mean field theory are also presented.
\end{abstract}

%
\maketitle
\section{\label{sec:intro}Introduction}
The search for suitable materials for spintronic application has spurred a large interest in materials that can be deposited on ordinary semiconductors such as Si and GaAs. Bulk MnAs exhibits a ferromagnetic (FM) transition temperature ($T_{c}$) of 313 K, \cite{Hilpert1911} where it is accompanied by a structural phase transition going from the FM $\alpha$-phase to the orthorhombic $\beta$-phase and as the temperature is raised to 398 K MnAs exhibits an additional phase transition back to a hexagonal $\gamma$-phase. The long term interest in MnAs lies in the first order magnetostructural phase transition with a $T_{c}$ that can be altered by for instance applying a stress to the material resulting in tensile/compressive strain of the MnAs lattice. \cite{Menyuk1969} When MnAs is deposited on a GaAs substrate as a thin film, MnAs undergoes a strain induced increase in $T_{c}$ and moreover develops a two phase region where the $\alpha$- and $\beta$-phases coexist between approximately 280 and 340 K.\cite{Das2003} The magnetic transition temperature depends on choice of substrate, e.g. MnAs/GaAs(001) has a $T_c$=320 K while MnAs/GaAs(111)B has a $T_c$=346 K, \cite{Wikberg2008} as well as on strain, e.g. stretching of the $c$-axis and compression of the  $a$-axis induce a reduction of $T_{c}$. \cite{Iikawa2005} Even though the films are fully relaxed after just a few monolayers \cite{Jenichen2004} the strain induced $T_{c}$ increase seems to prevail in thicker films and through application of high magnetic fields the $\beta$-phase can be transformed back to the $\alpha$-phase even above $T_{c}$ due to the structural strain. \cite{Ney2005} The magnetic nature of the $\beta$-phase has been under debate where some consider it to be paramagnetic \cite{Lindner2004} while later suggestions point out that it has an antiferromagnetic (AFM) short ranged structure, \cite{Yamaguchi2005} albeit without long range order. \cite{Niranjan2004} Investigations of the magnetic anisotropy have determined, counter intuitively, that for MnAs the easy axis of magnetization lies along the $a$-axis ([$11\overline{2}0$]-direction) and the hard axis along the $c$-axis ([0001]-direction). \cite{Schippan2000} Since both $c$ and $a$ lie in the film plane of MnAs/GaAs(001) films the magnetocrystalline anisotropy can be measured by simply rotating the film 90$^{\circ}$. An additional contribution to the in-plane anisotropy is shape-anisotropy arising from the striped structure with alternating $\alpha$- and $\beta$-stripes, stretching out along [0001], that decrease and grow in width, respectively, as the temperature increases. \cite{Ryu2008} Reported values of the anisotropy constant \cite{Lindner2004} are close to values obtained for bulk MnAs ($7.4\times10^{5}$ J/m$^{3}$ at room temperature).\cite{Blois1963}
\newline We present an investigation of MnAs films grown on GaAs(100) substrates and describe how the magnetic properties depend on strain and uniaxial structure of the thin film. The temperature dependent 1st and 2nd order terms ($K_{1}$ and $K_{2}$) of the magnetocrystalline anisotropy (MCA) are derived from the free energy density equation revealing an \textit{easy plane} anisotropy in MnAs films at all temperatures. X-ray magnetic circular dichroism (XMCD) measurements reveal that the orbital moment does not strictly follow the spin magnetic moment in the temperature range where both the $\alpha$- and $\beta$-phases coexist.

\section{\label{sec:Experi}Experimental} 
\begin{figure}[htp]
\includegraphics[angle=0,width=0.45\textwidth]{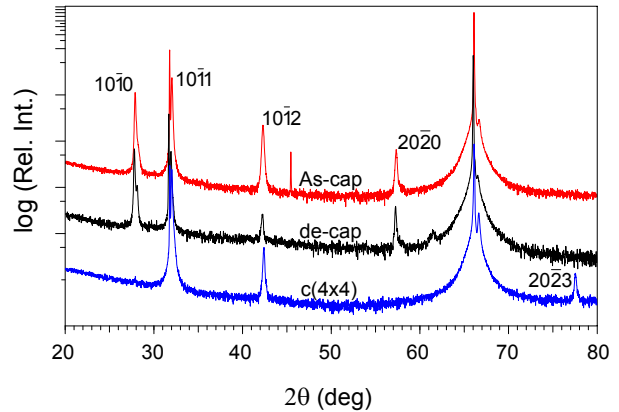}
\caption{\label{XRD} (Color online) X-ray diffraction $\theta$-2$\theta$ scans for the As-cap (red), de-cap (blue) and c($4 \times 4$) (black) samples obtained at 297 K. All samples show good epitaxy but with several orientations present in the films, see Table \ref{tab:XRD}. }
\end{figure}

The MnAs films were grown in a KRYOVAK MBE system, using epi-ready GaAs wafers with (100) orientation as substrates. The substrates were attached to molybdenum holders by liquid In gluing, which ensures uniformity of the substrate temperature during the MBE growth. After introduction into the MBE growth chamber, the substrates were subjected to the typical procedures of thermal evaporation of native oxides and high temperature growth of a GaAs buffer layer. An As cracking cell (by DCA Instruments) operated at 900$^\text{o}$ C was used as the arsenic source. After the buffer layer growth the temperature of the As cracker was lowered to 600$^\text{o}$C, which changes the As flux from As dimer rich to As tetramer (As$_{4}$) rich
and the substrate temperature was lowered to about 250 $^\text{o}$C. This temperature was measured by an IR pyrometer operating in the 100 - 700$^\text{o}$C range. During the cooling time the substrates were exposed to the As$_{4}$ flux. The desired surface 
reconstruction (c(4$\times$4) or d(4$\times$4)) of the GaAs substrate was obtained  by following established procedures. \cite{Jiricek2009,substrate} The MnAs growth was started after reaching and stabilizing the substrate temperature at 250$^\text{o}$C; the MnAs growth rate was chosen to be about 200{\AA}$/$h. The growth was monitored by a reflection energy electron diffraction (RHEED) system. The RHEED images evolved rather quickly from two-dimensional (2D) 4$\times$4 streaky patterns originating from the GaAs surface to diffraction spots indicating formation of MnAs islands during the initial growth stage. When the growth proceeded the RHEED images from MnAs:GaAs(100) evolved to an admixture of three-dimensional spotty and 2D streaky patterns, which were visible up to a MnAs film thickness of about 500 {\AA}. The spotty patterns eventually disappeared for thicker MnAs films. To prevent MnAs surface oxidation after the MBE growth the samples were capped with a 1000 {\AA} thick amorphous As layer deposited in the MBE chamber after switching off the substrate heater and cooling down the sample to the lowest temperature possible to attain in the MBE system. 
\newline One of the investigated samples had a c(4$\times$4) surface reconstruction of the GaAs(100) surface prior to MnAs growth, henceforth referred to as the c(4$\times$4) sample, while the As- and de-cap samples had a d(4$\times$4) surface reconstruction of the GaAs(100) surface prior to MnAs growth. The de-cap sample was subjected to an additional post heat treatment at 310$^\circ$C for 60 minutes (taking place during removal of the As capping layer) in vacuum. 
\newline X-ray diffraction (XRD) $\theta - 2\theta$ scans were carried out on a Bruker D8 Bragg-Brentano system, equipped with a $\mathrm{Cu}\,K_{\alpha_1}$ monochromator and a V{\aa}ntec detector. Pole figures were measured using a Phillips MRD system with point focus and an X-ray poly capillary lens and a secondary 0.18$^{\circ}$ parallel plate collimator and a graphite monochromator. This set-up was also used to measure the $c$-axis (($0004$) peaks) of the MnAs ($10\overline{1}0$) orientation using grazing incidence XRD, with an incoming angle of about 0.7-0.8$^{\circ}$. For the MnAs ($10\overline{1}1$) orientation an asymmetric scan was used to measure the $c$-axis. The MRD system was also used for rocking curve measurements, using a 2-bounce hybrid monochromator/mirror and secondary X-ray mirror set-up. Due to the difference in scattering power of the ($10\overline{1}0$), ($10\overline{1}1$) and ($10\overline{1}2$) planes the peak intensities in the $\theta$-$2\theta$ diffractogram shown in Fig.~ \ref{XRD} are not proportional to the amount of each orientation. The ($10\overline{1}1$) orientation is overrepresented due the high scattering factor of this plane (about 40 $\times$ higher than for the ($10\overline{1}0$) plane). The proportion of each orientation was therefore estimated by comparing the intensities in ($11\overline{2}0$) pole figures, cf. Fig.~\ref{polefig}.
\newline A transmission electron microscopy (TEM) study was performed as a function of temperature between 89 and 325K on a d(4$\times$4) surface reconstructed sample using a JEOL-2100 microscope operated at 200keV with a double-tilt cooling holder (Gatan CHDT3504). The cross-section TEM specimen was obtained, after mechanical lapping and polishing, by ion-milling (JEOL Ion-slicer) thinning. The selected area electron diffraction (SAED)  patterns along [101] GaAs (or [$10\overline{1}0$] MnAs), which were recorded on a CCD camera (Gatan ES500W), were used to determine the $a$- and $c$-axis at different temperatures. As a reference, the (0-20) GaAs reflection was also measured. Two selected areas of the sample, each with an effective size of 800 nm in diameter, were investigated but no significant difference in the results were found between the studied areas.
\newline X-ray magnetic circular dichroism (XMCD) measurements were performed at beamline D1011 at the MAX-lab synchrotron facility.\cite{lund1} All measurements were performed by magnetizing the sample with an in-plane magnetic field in the easy magnetization direction after which the absorption spectrum was measured in remanence using total electron yield. The measurements were performed using 75$\pm$5\% circularly polarized light and with the sample 13$^{\circ}$ from grazing incidence. The same procedure was repeated after magnetizing the sample in the opposite direction. Saturation effects which are prominent at low angles are compensated for according to Nakajima \textit{et al.}\cite{Nakajima} The spectra are normalized at low photon energies after which a step function is subtracted. We have used a correction factor for the spin magnetic moment of 1.47, which has been suggested to remedy problems arising when attempting to separate overlapping contributions (due to the relatively small spin-orbit splitting of the Mn $L$ shell levels) from the $L_3$ and $L_2$ edges.\cite{Edmonds}
\newline Magnetization measurements were performed in a Quantum Design MPMS-XL Superconducting Quantum Interference Device (SQUID) magnetometer. Magnetization ($M$) versus temperature ($T$) was studied between 5 K and 400 K following a field cooled protocol. A weak magnetic field ($H = 50$ Oe) was applied along the crystallographic [$11\overline{2}0$], the easy axis of magnetization, at 400 K and the magnetization was measured as the sample cooled down to 5 K. Magnetization versus field measurements were performed at different temperatures in the range 10 - 340 K between 50 kOe and -50 kOe. The field was applied either along [$11\overline{2}0$] or along  [$0001$], the in-plane easy and hard axis of magnetization, respectively.
\newline To corroborate the experimental studies we performed theoretical investigations based on density functional theory. In these studies, we have used linearized muffin tin orbital (LMTO) calculations with no shape approximation for charge density or potential.\cite{Wills} The lattice is described by a non-overlapping spherical region surrounding atomic sites called muffin-tin sphere and in between them an interstitial part. The basis set is formed with augmented linear muffin-tin orbitals. In-side the muffin-tin sphere basis function, charge density and potential are expanded in spherical harmonic functions up to $l_{max}=8$ along with radial functions and in a Fourier series in the interstitial region. In the interstitial region the basis functions are Bloch sums of spherical Hankel and Neumann functions. The Dirac equation is solved for a core charge density which means that no frozen core approximation is considered. A multi basis formalism is used to ensure that all wave functions are well converged, i.e. we have considered three 4s, three 4p and two 3d orbitals for Mn and three 4s, three 4p and two 4d orbitals for As in the wave function. The modified tetrahedron method is used for integration in reciprocal space over 1152 k-points in the whole Brillouin zone (BZ) for self-consistent ground state calculations and to ensure convergence of the magnetic anisotropy energy (MAE), we carried out up to 9216 k-points in the full BZ. The exchange correlation potential is approximated by the PBE96 functional within the generalized gradient approximation. We also performed studies of the influence of electron correlations using dynamical mean field theory.

\section{\label{sec:Results}Results}
\subsection{\label{sec:XRD}Structural properties}
Several out-of-plane and in-plane orientations have been reported for MnAs films deposited on GaAs (100) substrates. These orientations have been classified as $A$ or $B$ depending on the in-plane relationship with respect to the substrate. \cite{Iikawa2004, Tanaka1, Tanaka2, Tanaka3} The $A$ and $B$ orientations have a twin relationship of $90^\text{o}$ where in the $A$-type the $c$-axis of MnAs is parallel with the GaAs [$1\overline{1}0$] direction and for the $B$-type the MnAs $c$-axis is parallel to the GaAs [110] direction. Moreover, an index in the notation for the orientation was introduced to distinguish between in-plane and out-of-plane orientations of the MnAs $c$-axis. \cite{Daweritz1, Daweritz2} For $A_{0}$ and $B_{0}$ there is an in-plane orientation of the MnAs $c$-axis, while for $A_{1}$, $B_{1}$ and $A_{2}$, $B_{2}$ the $c$-axis is tilted out-of-plane with respect to the GaAs(001) surface (see Table \ref{tab:XRD} for details). The orientation of the MnAs film can be controlled by the template layer. More in particular, a very As-rich template with a d(4$\times$4) surface reconstruction yields an $A_{0}$ orientation while reduction of the As-concentration on the template yields a $B_{0}$ or even a $B_{1}$ orientation. \cite{Iikawa2004} 

\begin{figure}[htp]
\includegraphics[angle=0,width=0.45\textwidth]{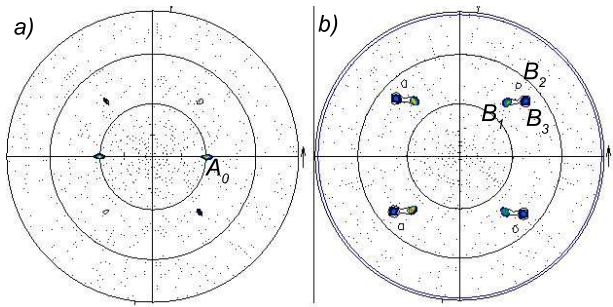}
\caption{\label{polefig} $11\overline{2}0$ pole figures in linear scale with indications of orientations ($A_{0}$, $B_{1}$, $B_{2}$ and $B_{3}$) for {\bf a)} the de-cap sample with mostly $A_{0}$ orientation and a very small $B_{1}$ contribution and for  {\bf b)} the c(4$\times$4) sample with clear $B_{1}$ and $B_{3}$ orientation peaks.}
\end{figure}

The d(4$\times$4) films show a dominating ($10\overline{1}0$) orientation ($A_{0}$). Small amounts of additional ($10\overline{1}1$) and ($10\overline{1}2$) out-of-plane orientations were also observed. The fraction of ($10\overline{1}1$) orientation was estimated to about 5-13 \%, while the ($10\overline{1}2$) orientation was less abundant, 0.4-2\%. The de-cap film shows less ($10\overline{1}1$) and ($10\overline{1}2$)  out-of-plane orientations, probably due to the heat treatment during the As-layer removal. The $A_{0}$ orientation exhibits a rocking curve Full Width Half Maximum (FWHM) value of about 0.5$^{\circ}$ for the de-cap sample and 0.65$^{\circ}$ for the As-cap sample, while the $B_{1}$ orientations have rocking curve FWHM values of 0.4-0.5$^{\circ}$. The $B_{2}$ orientation shows an even higher FWHM value, 1.2$^{\circ}$ and 1.9$^{\circ}$ for the As-cap and de-cap samples, respectively. These relatively high rocking curve FWHM values indicate a high degree of strain relaxation in the $a$- and $b$-axis directions.
\newline The c(4$\times$4) film exhibits a dominating ($10\overline{1}1$) orientation with smaller amounts of the ($10\overline{1}2$)  and ($20\overline{2}3$) orientations, 2\% and 25\%, respectively. The rocking curve FWHM value for the dominating orientation is about 0.5$^{\circ}$, while the ($10\overline{1}2$) orientation exhibits a higher value, about 1.0$^{\circ}$. The ($20\overline{2}3$)  orientation exhibits two rocking curve peaks, $\pm$1.7$^{\circ}$ from the $\theta$ angle. The FWHM value of the ($20\overline{2}3$)  rocking curve peak is about 1.5-1.6$^{\circ}$. This indicates that the fraction of ($20\overline{2}3$) orientation could be underestimated due to the high FWHM value. The films are almost completely relaxed with $a$ exhibiting values close to the bulk value, while a strain up to $1\%$ is observed for $c$ (see Table \ref{tab:anisotropy}). In order to investigate the in-plane orientation and the fraction of each orientation present in the films, pole figures were measured. From the  ($11\overline{2}0$)~pole figure, shown in Fig.~\ref{polefig} for the de-cap and c(4$\times$4) samples, the relative fraction (see Table \ref{tab:anisotropy}) of each orientation could be estimated through the height of each peak. It should be noted that $A/B$-orientations with the same $\alpha$ will appear on the same $\psi$-angle in the ($11\overline{2}0$)~pole figure.

\begin{table}
\caption{\label{tab:XRD} Epitaxial orientations of MnAs films deposited on GaAs (100), using the  notation defined in Refs. \onlinecite{Tanaka1,Tanaka2,Tanaka3,Ozeki}, as well as the corresponding  out-of-plane angles ($\alpha$) of the $c$-axis with respect to the GaAs (001) surface.}
\begin{ruledtabular}
\begin{tabular} {cccc}
Orientation (hkl) & Miller-Bravais & $\alpha$ & Notation \\ 
\hline
(100) & $10\overline{1}0$ & $0^\circ$ & $A_{0}$, $B_{0}$ \\
(101) & $10\overline{1}1$ & $29^\circ$ & $A_{1}$, $B_{1}$ \\
(102) & $10\overline{1}2$ & $41^\circ$ & $A_{2}$, $B_{2}$ \\
(203) & $20\overline{2}3$ & $50^\circ$ & $A_{3}$, $B_{3}$ \\ 
\end{tabular}
\end{ruledtabular}
\end{table}

In the c(4$\times$4) sample, the out-of-plane orientations ($B_{1}$ and $B_{3}$) dominate and almost no in-plane orientations of $c$ ($A_{0}$/$B_{0}$) with respect to the substrate surface are found. For the d(4$\times$4) surface reconstructed samples the $A_{0}$ orientation dominates and through heat treatment a relaxation of the $c$-axis occurs (0.3\%) together with an increase of the fraction of $A_{0}$ orientation. Both the de-cap and As-cap samples have a fraction of $B$-orientation, which is responsible for the small but finite value of the remanent magnetization when measuring the magnetization with the field applied along the hard magnetization direction, cf. Fig. \ref{MvsHall}. The out-of-plane $A$-orientations (A$_{1}$ and A$_{2}$) will also decrease the remanent magnetization when applying the field along the easy axis of magnetization (to be discussed below).
   
\begin{figure}[htp]
\includegraphics[angle=0,width=0.45\textwidth]{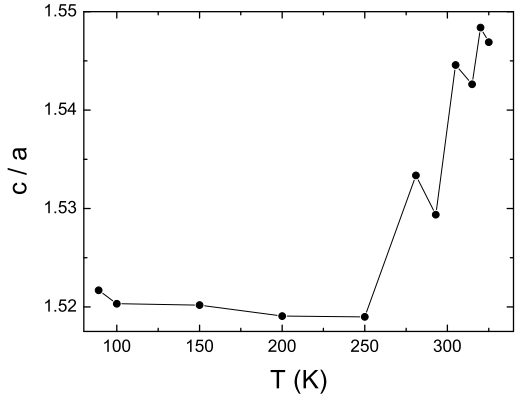}
\caption{\label{SAED} Temperature dependence of the $c/a$-ratio. The $c$- and $a$-axis data were extracted from SAED patterns.}
\end{figure}

Figure~\ref{SAED} shows the temperature dependence of the $c/a$-ratio as derived from the SAED patterns. Even though the SAED patterns are dominated by the $\alpha$-phase reflections, some weak reflections due to the $\beta$-phase can be resolved at the highest temperatures. The $c/a$-ratio starts to increase at about 250 K, which is where we expect the low temperature limit of the $\alpha - \beta$ coexistence region to be (see below), and reaches a value close to 1.55 at the highest temperature. This increase of the $c/a$-ratio is due to a 1\% increase of $c$ and a somewhat smaller decrease (0.7\%) of $a$. The results also indicate that the increase of $c$ starts at slightly lower temperature compared to the decrease of $a$. 
 
\subsection{\label{sec:Magn}Magnetic properties}
\begin{figure}[htp]
\includegraphics[angle=0,width=0.5\textwidth]{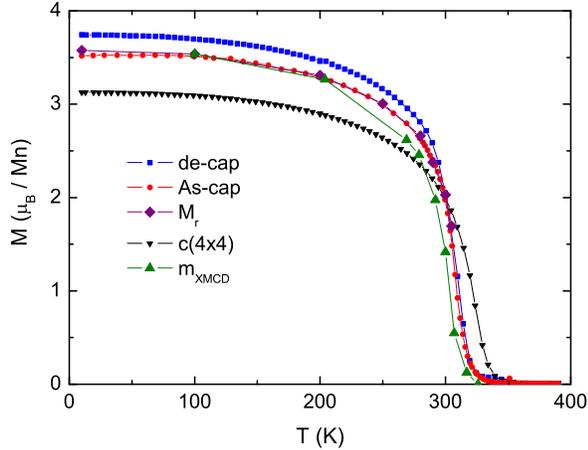}
\caption{\label{MvsT} (Color online) Magnetization ($M$) versus temperature ($T$) for the de-cap (blue squares), As-cap (red circles),  and c($4 \times 4$) (black triangles) samples. The remanent magnetization ($M_r$, purple diamonds) for the de-cap sample and the magnetic moment ($m_\mathrm{XMCD}$, green triangles) obtained from XMCD measurements are included for comparison. The $m_\mathrm{XMCD}$ data have been normalized with respect to $M_r$ at 100 K. }
\end{figure}

\begin{figure}[htp]
\includegraphics[angle=0,width=0.5\textwidth]{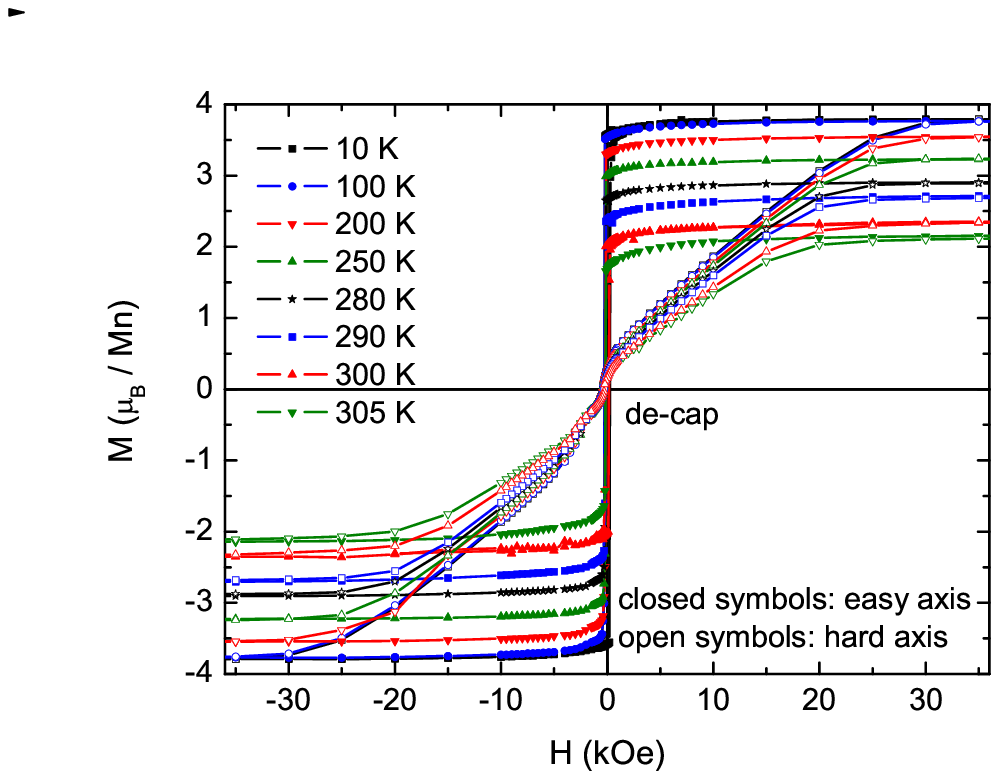}
\caption{\label{MvsHno} (Color online) Magnetization ($M$) versus field ($H$) at different temperatures in the range $10-305$ K, for the  de-cap sample with filled symbols corresponding to field applied along the easy axis ([$11\overline{2}0$]-direction) of magnetization and open symbols to field applied along the hard axis ([0001]-direction) of magnetization. }
\end{figure}

From the temperature dependent magnetization curves shown in Fig. \ref{MvsT} it is seen that the highest $T_{c}$ is obtained for the c(4$\times$4) sample followed by the de-cap and As-cap samples (see Table \ref{tab:anisotropy} for values). For comparison, the temperature dependence of the remanent magnetization for the de-cap sample ($M_r$) along with the magnetic moment obtained from XMCD measurements ($m_\mathrm{XMCD}$) normalized with respect to $M_{r}$ are plotted in Fig.~\ref{MvsT}. The archetypical behavior of the field dependent magnetization in MnAs/GaAs(100) thin films at different temperatures and fields applied along [$11\overline{2}0$] and [$0001$], here represented by the results obtained for the de-cap sample, is seen in Fig. \ref{MvsHno}. When measured along the easy axis of magnetization (closed symbols) the film shows a very square like and narrow hysteresis curve, with a coercivity around 160 Oe at 10 K. The hard axis magnetization exhibits a nearly linear relationship with the applied field up to saturation with a saturating field of $\approx 35$ kOe at 10 K. That the hard axis magnetization exhibits a small but finite remanent magnetization as well as magnetic hysteresis is related to the small fraction of out-of-plane $B$ orientation in the film, as seen in the XRD-measurements, i.e. a small part of the film will have easy axis or close to easy axis alignment.


MnAs has its easy axis of magnetization along the $a$-axis ([$11\overline{2}0$]-direction) and the hard axis along the $c$-axis ([0001]-direction) in the hexagonal unit cell. When  MnAs is deposited on GaAs(100) both the $a$- and $c$-axis will be oriented in the film plane and thus the magnetic anisotropy can be studied by performing $M$ versus $H$ measurements along  the in-plane directions defined by these two axes. In case of uniaxial crystal structure, the expression for the magnetic free energy density (\textit{F}) is given by: 
\begin{eqnarray}
F & = & K_{1}[a_{0}\cos^{2}\alpha_{0} + a_{1}\cos^{2}\alpha_{1} + \dots]\sin^{2}\theta \,+ {}\nonumber\\
&& +\, K_{2}[a_{0}\cos^{2}\alpha_{0} + a_{1}\cos^{2}\alpha_{1} + \dots]\sin^{4}\theta \,- {}\nonumber\\
&& -\, \mu_{o}HM_{s}\cos\theta ,
\label{E_equation}
\end{eqnarray}
where $K_{1}$ and $K_{2}$ are the magnetic anisotropy constants, $a_{0}$, $a_{1}$$\dots$ are the fractions of $A_{0}$, $A_{1}$$\dots$ orientations, respectively, in the sample and $\alpha_{0}$, $\alpha_{1}$$\dots$ are the corresponding out-of-plane angles of the $c$-axis with respect to the film plane. The last term is the Zeeman contribution to the free energy and $\theta$ is the angle between the magnetization and applied field directions. Through the relationships $M=M_{s}cos\theta$ and $\frac{\partial}{\partial\theta} F=0$ an expression for field dependence of the hard axis magnetization, $M^{hard}/M_{s}$ vs. $H$ can be derived, which when fitted to the corresponding experimental data yields the anisotropy constants. The values for $K_{1}$ and $K_{2}$, given in Table \ref{tab:anisotropy}, have been obtained by subtracting the easy-axis magnetization contribution in the $M^{hard}$ vs. $H$ data before fitting the $M^{hard}/M_{s}$ expression to the experimental data. As seen in Table \ref{tab:anisotropy} the As-cap and de-cap samples show a somewhat higher anisotropy and $T_c$ than obtained for bulk MnAs, \cite{Blois1963} while the $c(4\times4)$ sample displays lower $K$-values but a higher $T_c$ value.   

\begin{table*}
  \caption{\label{tab:anisotropy} Anisotropy constants $K_{1}$ and $K_{2}$ measured at different temperatures ($T_{m}$), $T_{c}$, room temperature values for $c$-axis length and $c/a$-ratio, and fraction of each in- and out-of-plane orientation for the different samples of 1000 \AA~thick MnAs/GaAs(100), compared with bulk values from Ref.~\onlinecite{Blois1963}.}
\begin{ruledtabular}
\begin{tabular}{cccccccc} 
\mbox{sample} & \mbox{$T_{m}$ (K)} & \mbox{$K_{1} ( \times 10^{5}$ J/m$^{3}$)} & \mbox{$K_{2} ( \times 10^{5}$ J/m$^{3}$)} & \mbox{$T_{c}$ (K)} & \mbox{$c_\mathrm{hex}$ (\AA)} & $c/a$ & \mbox{A$_{0}$~/~B$_{1}$~/~A$_{2}$~/~B$_{2}$~/~B$_3$} \\ 

\hline
As-cap & 10 & -14.0 & 1.4 & 317 & 5.772 & 1.555 & 0.86~/~0.13~/~0.017~/ - / - \\
 & 100 & -13.9 & 1.3 &  &  &  \\
 & 200 & -12.4 & 1.2 &  &  &  \\
 & 250 & -10.3 & 0.9 &  &  &  \\
 & 280 & -8.4 & 0.9 &  &  &  \\
 & 290 & -7.5 & 0.9 &  &  &  \\
 & 300 & -6.2 & 0.6 &  &  &  \\
 & 305 & -5.9 & 0.4 &  &  &  \\
\hline
de-cap & 10 & -14.4 & 1.7 & 320 & 5.755 & 1.551 & 0.95~/~0.050~/ - /~0.004~/ - \\
 & 100 & -14.4 & 1.7 &  &  &  \\
 & 200 & -12.8 & 1.5 &  &  &  \\
 & 250 & -10.6 & 1.2 &  &  &  \\
 & 280 & -8.8 & 1.1 &  &  &  \\
 & 290 & -7.9 & 1.1 &  &  &  \\
 & 300 & -6.5 & 0.8 &  &  &  \\
 & 305 & -5.8 & 0.6 &  &  &  \\
 \hline
c(4$\times$4) & 10 & -7.3 & 1.5 & 335 & 5.746 & 1.544 & 0.008~/~0.72~/ - /~0.024~/~0.25 \\
 & 100 & -5.6 & 1.1 &  &  &  \\
 & 200 & -4.8 & 0.9 &  &  &  \\
 & 300 & -3.2 & 0.8 &  &  &  \\
 & 320 & -2.9 & 0.9 &  &  &  \\
\hline
Bulk & 77 & -12.0 & - &  313 & 5.712 & 1.540 & \\
 & 198 & -11.0 & - &  &  &  \\
 & 273 & -9.0 & - &  &  &  \\ 
 & 299 & -7.6 & - &  &  &  \\
\end{tabular}
\end{ruledtabular}
\end{table*}

\begin{figure}[htp]
\includegraphics[angle=0,width=0.45\textwidth]{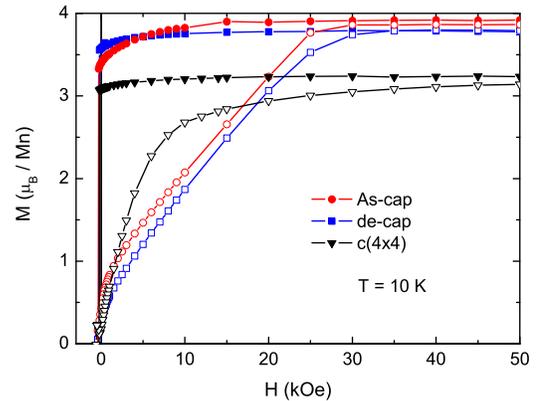}
\caption{\label{MvsHall} Magnetization ($M$) versus field ($H$) measured at 10 K along the easy axis (filled symbols) and hard axis (open symbols) of magnetization, with the As-cap sample in red circles, de-cap sample in blue squares and c(4$\times$4) sample in black triangles. }
\end{figure}

A difference in magnetic behavior can also be seen in the field dependence of the magnetization, see Fig.~\ref{MvsHall} where the closed and open symbols correspond to measurements along [$11\overline{2}0$] and [0001], respectively. A clear difference in the values for the saturation magnetization ($M_{s}$) is seen between the samples. We believe that the differences in $T_{c}$, $M_{s}$ and MCA between samples are strain related and connected to the existence of multi-epitaxial orientations in the films. The c($4 \times 4$) sample exhibits a dominant out-of-plane orientation that is responsible for the more rounded shape of the $M^{hard}$ vs. $H$ curve as well as for the higher $T_{c}$. \cite{Iikawa2004} Even though the out-of-plane angles are compensated for when deriving $K_{1}$ and $K_{2}$, the obtained $K_{1}$ values are a factor of 2 smaller than the values obtained for the As-cap and de-cap samples as well as values derived for bulk samples.\cite{Blois1963} It should be noted that the easy-axis magnetization measurements on the c(4$\times$4) sample are made with the field applied along GaAs[1$\overline{1}$0] instead of GaAs[110]. Although the As-cap and de-cap samples have a more uniaxial orientation with 86$\%$ and 92$\%$ of $A_{0}$-orientation, respectively, they still show some dissimilarities in $T_{c}$, $M_{s}$ as well as in $K_{1}$ and $K_{2}$. This clearly corresponds well to the early investigations on bulk MnAs describing changes in magnetic properties with strain, \cite{Menyuk1969} whereas in the present study the unit cell distortion comes from lattice mismatch and multi-orientation induced strain. The increase in $T_{c}$ with the relaxing $c$ is well in line with reports from Iikawa \textit{et al.}, \cite{Iikawa2005} who showed a relationship between compression/stretch of the lattice and changes in $T_c$. We would like to point out that the variations in $T_c$ are due to two effects; the fraction and angle ($\alpha$) of $c$-axis out-of-plane orientations \cite{Iikawa2004} and the stretching/relaxation of the $c$-axis. The interdependence between $\alpha$ and $T_{c}$ is evident for the c(4$\times$4) sample where no in-plane alignment is detected. Although the c(4$\times$4) sample has a close to bulk $c/a$-ratio (see Table \ref{tab:anisotropy}) the anisotropy seems to be more dependent on $c$-axis alignment along GaAs[1$\overline{1}$0] than on $c$-axis strain. A conclusion that becomes even more evident when examining the $c/a$-ratio for the As-cap and de-cap samples.   

\subsubsection{X-ray Magnetic Circular Dichroism}
\begin{figure}
     \includegraphics[width=0.450\textwidth]{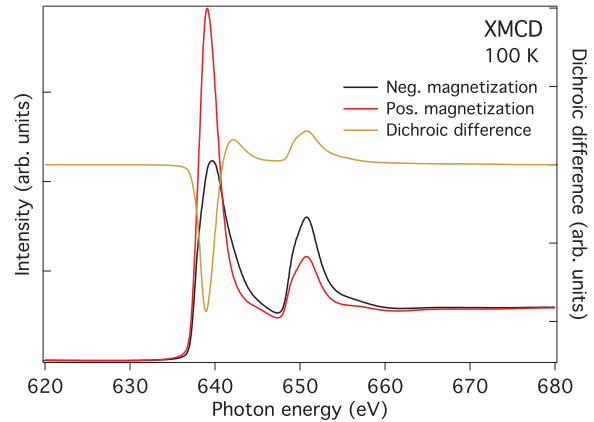}
     \caption{\label{XMCD_100K} Mn L$_{2,3}$ XMCD spectra recorded at 100 K and their corresponding difference spectrum. Spectra were obtained by reversing the magnetization direction in the sample. }
\end{figure}

\begin{figure}
     \includegraphics[width=0.45\textwidth]{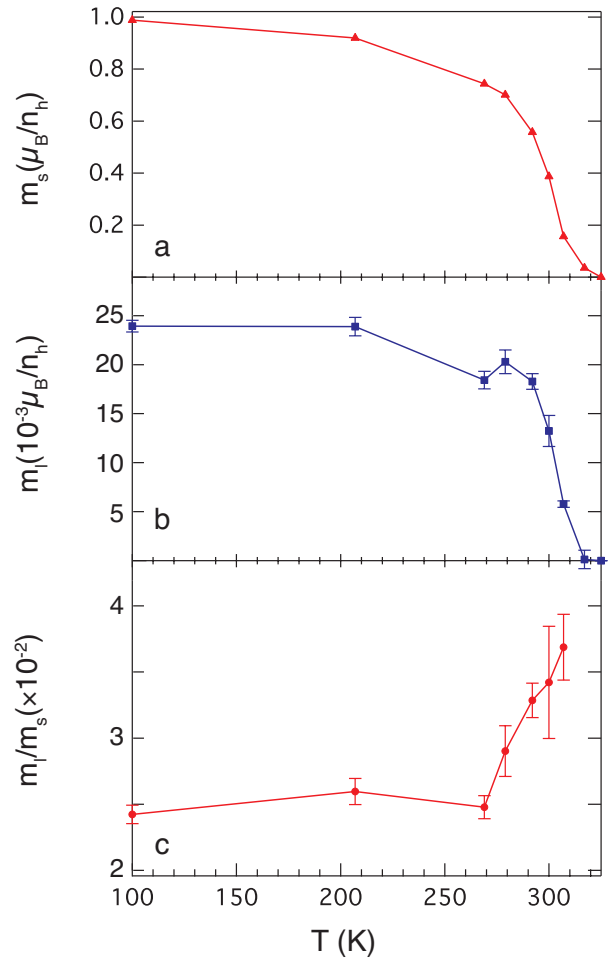}
     \caption{\label{XMCD_MvsT} \textbf{a)} Spin magnetic moment ($m_s$), \textbf{b)} orbital magnetic moment ($m_l$) and \textbf{c)} the ratio $m_l/m_s$.}
\end{figure}

Mn $L_{2,3}$ x-ray magnetic circular dichroism spectra recorded at 100~K for the de-cap sample are shown in Fig.~\ref{XMCD_100K} together with the corresponding dichroic difference. The spectra were obtained by reversing the magnetization of the sample keeping the polarization fixed. Similar spectra were recorded for several temperatures in the interval 100-330~K. The spectra have been normalized to a common edge jump corresponding to a per atom normalization of the absorption cross section.\cite{NEXAFSBookByStohr} The use of the XMCD sum rules \cite{Carra1993} for Mn has been controversial since there is a mixing between the $L_2$ and $L_3$ edges and therefore a constant of 1.47 has been proposed to compensate for this effect.\cite{Edmonds} Using a $d$-hole count of 4.9 as the number of empty $d$ states in the sum rules, a magnetic moment of $\sim$5 $\mu_B$/Mn atom is obtained at low temperatures. This number is clearly too large as compared to the numbers obtained from magnetization measurements and theory. This discrepancy might have several explanations. We first note that the spin-quadrupole term $\left\langle T_z \right\rangle$ was neglected in the sum rule analysis. A $d^5$ configuration would in a localized picture give the same occupation of all orbital levels resulting in a zero $\left\langle T_z \right\rangle$ term.\cite{Wu1994} However, one could of course argue that this approach might not be strictly valid for the MnAs system due hybridization effects and a deviation from a strict $d^5$ configuration. From our calculations we find that the $d$ occupation is 5.13 electrons. We have therefore also made explicit calculations of the $\left\langle T_z \right\rangle$ correction, following the recipe by Wu and Freeman.\cite{Wu1994} In no case we find that the correction is greater than $0.18~\mu_B$, i.e., a correction less than 5 percent of the total moment. We also note that without the corrective factor suggested in Ref.~\onlinecite{Edmonds}, we obtain a spin moment that is very close to the value obtained from calculations. Errors in the estimate of the number of holes, obtained from theory, can safely be ruled out, as this would require a charge transfer of almost 1.5 electrons not accounted for by the theoretical calculations. A more likely explanation for the discrepancy in magnetic moments connects to the value used for the correction factor and a more thorough investigation on the applicability of the corrective factor for the XMCD spin sum rule for MnAs is under preparation.\cite{Sanyal-unpublished} To avoid distracting the focus from the more interesting point regarding the behavior of the spin and orbital moments as a function of temperature, we have chosen to present moments obtained from XMCD sum rule analysis as magnetic moments per hole. The results are summarized in Fig.~\ref{XMCD_MvsT}. As shown in Fig.~\ref{XMCD_MvsT}b the orbital moment ($m_l$) behaves  different from the spin magnetic moment ($m_s$) given in Fig.~\ref{XMCD_MvsT}a. As most evident from the $m_l$/$m_s$-ratio, seen in Fig.~\ref{XMCD_MvsT}c, the orbital moment decreases more slowly than the spin magnetic moment in the temperature range where the $\alpha$- and $\beta$-phases coexist. This behavior is not strictly excluded by any physical principle as we are dealing with a structurally driven first order magnetic transition. We note however that the result is an indication of structural changes in the $\alpha$-phase in the coexistence region. This conjecture is confirmed by the temperature dependence of $c/a$ shown in Fig.~\ref{SAED}; the very similar temperature dependence of $c/a$ and $m_l / m_s$ is striking. Moreover, the observed variation of the orbital magnetic moment is in accordance with the theoretical calculations, discussed later in the text and seen in Fig.~\ref{maexch}, where it is shown that an increased $c/a$-ratio can yield a larger orbital moment anisotropy. 
 
\subsection{Theoretical Calculations}
\begin{figure}[ht]
\begin{center}
\includegraphics[width=0.45\textwidth]{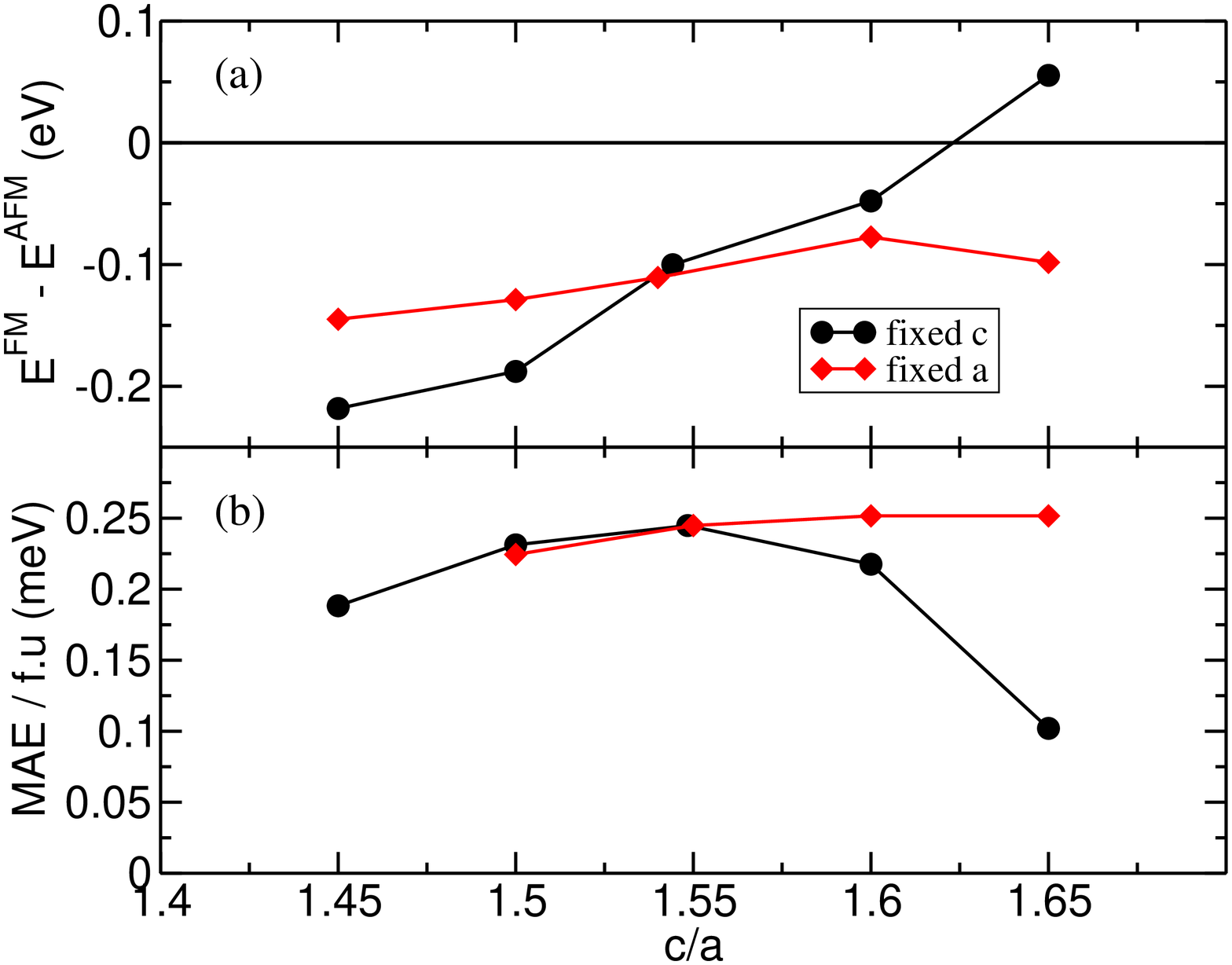}
\includegraphics[width=0.45\textwidth]{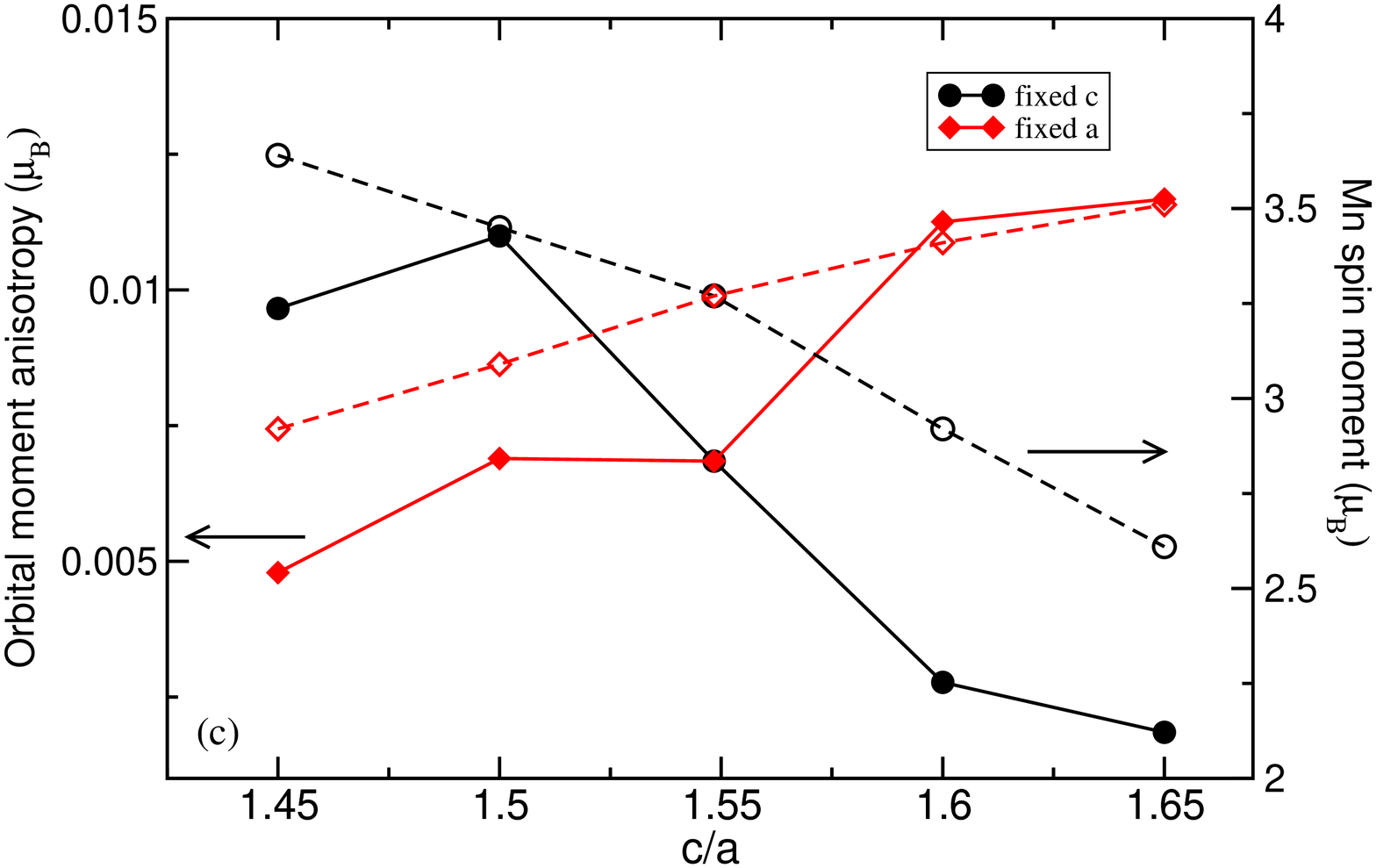}
\caption{\label{maexch} (Color online) Calculations for different $c/a$-ratio, where black symbols (red symbols) correspond to keeping the $c$-axis ($a$-axis) fixed while varying the $a$-axis ($c$-axis). {\bf a)} difference in total energies between ferromagnetic and antiferromagnetic alignments of Mn spins in the $\left[ 0001 \right]$-direction. {\bf b)} magnetic anisotropy energy (MAE) as a function of $c/a$-ratio, {\bf c)} orbital moment anisotropy (OMA) as a function of $c/a$-ratio on the left hand axis with filled symbols and solid line and spin magnetic moment as a function of $c/a$-ratio on the right hand axis with open symbols and dashed line. } 
\end{center}
\end{figure}

\begin{figure}[h!]
\begin{center}
\includegraphics[width=0.4\textwidth]{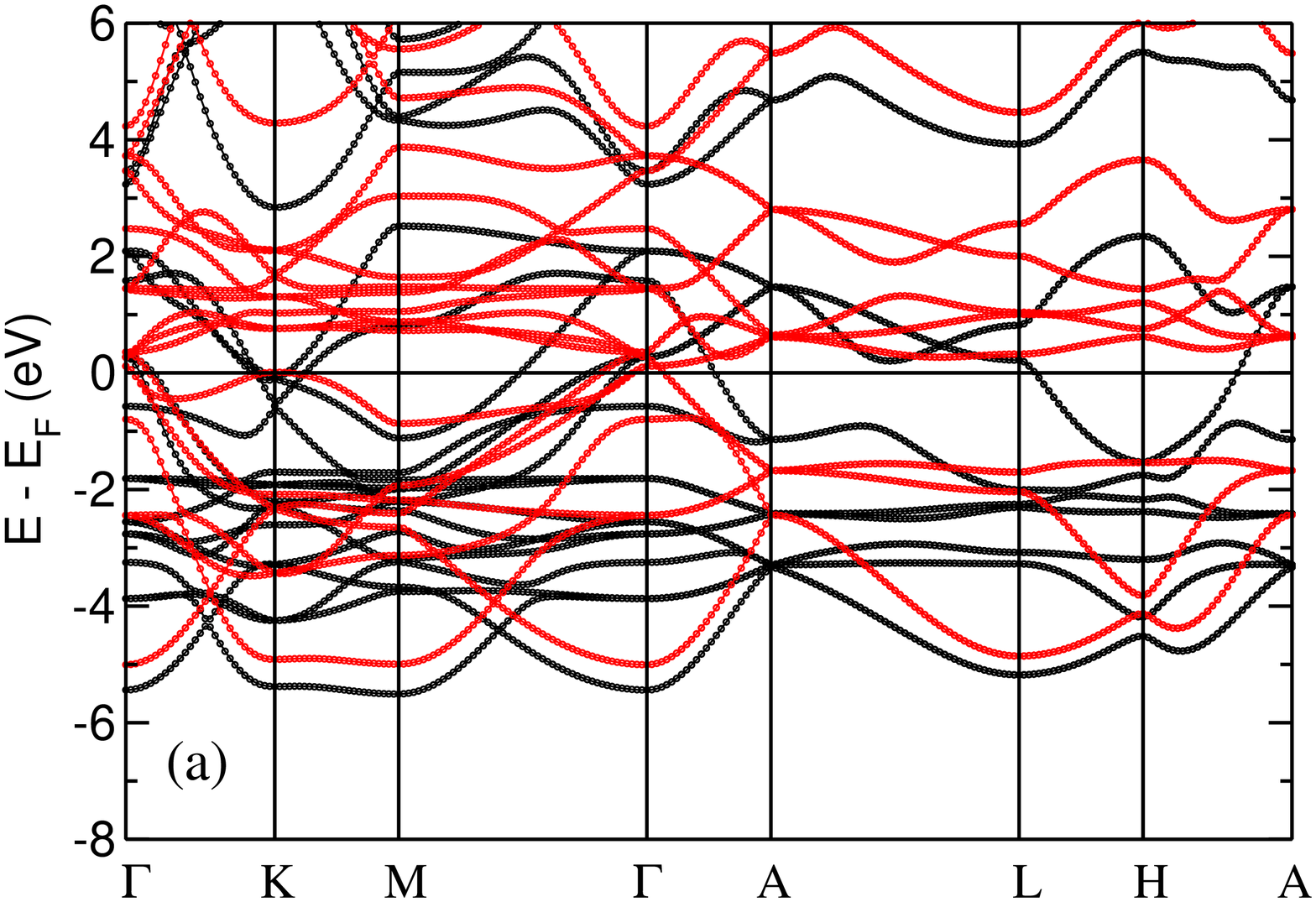}
\includegraphics[width=0.4\textwidth]{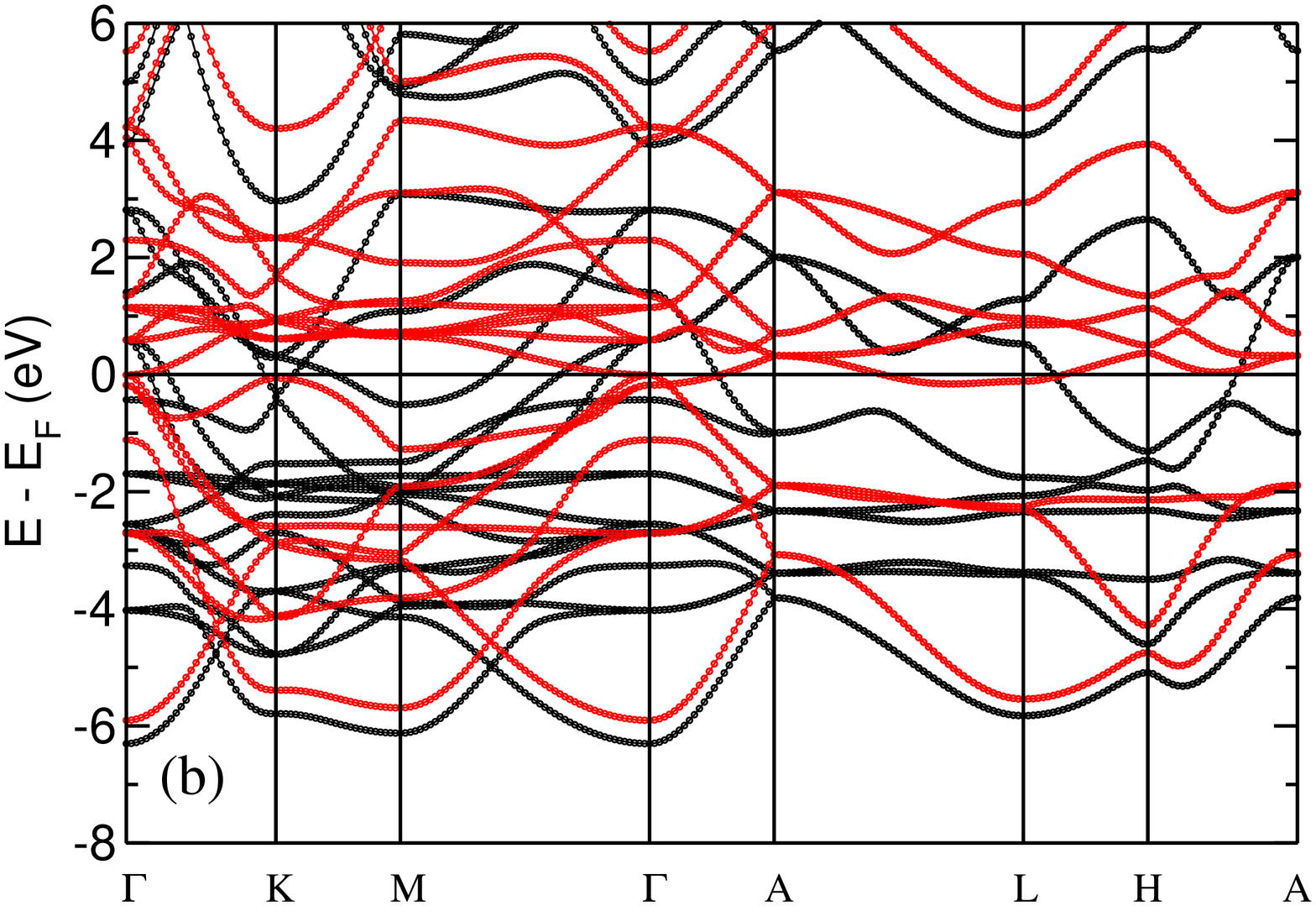}
\includegraphics[width=0.4\textwidth]{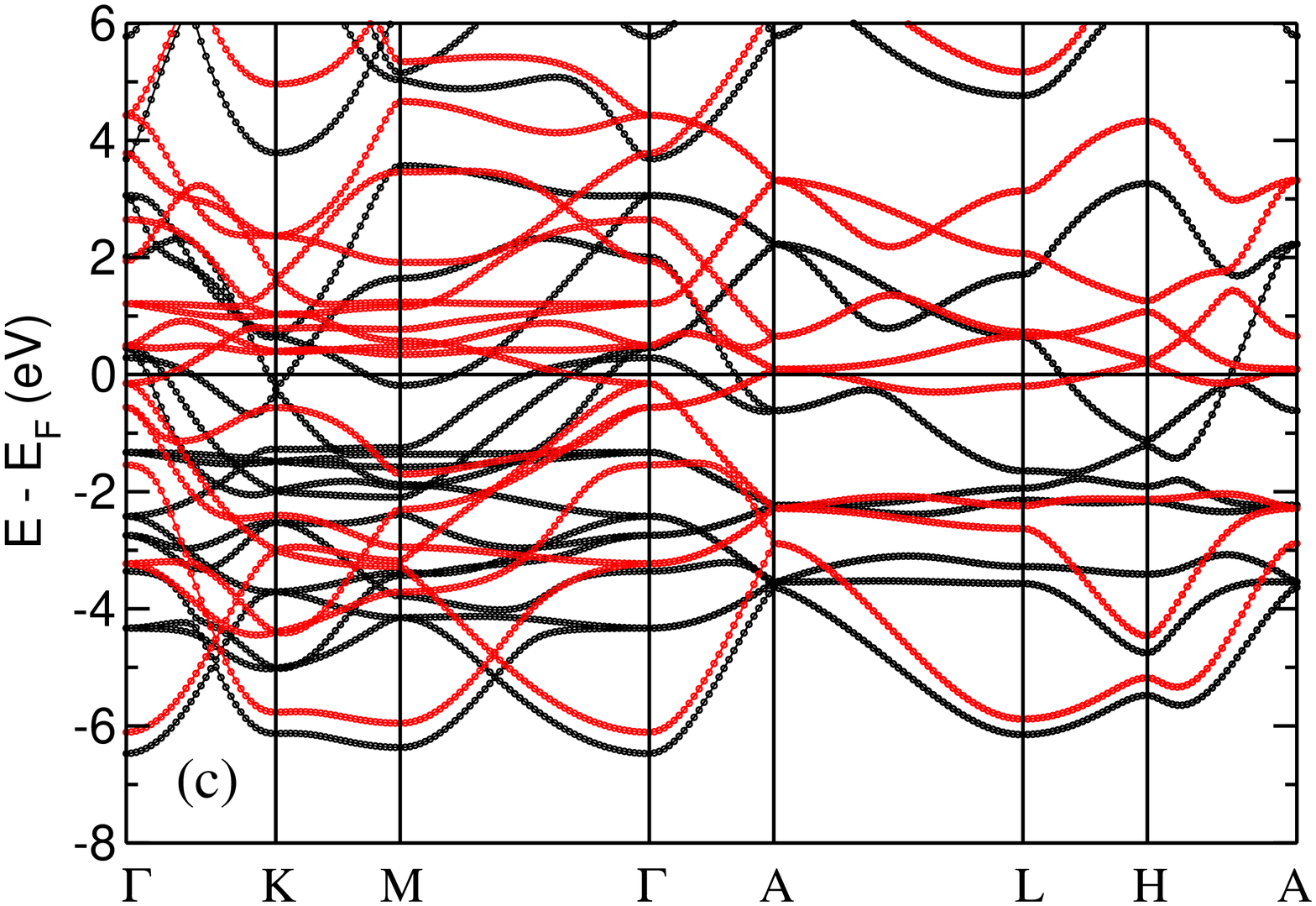}
\caption{(Color online) Spin-polarized band structures where black and red symbols indicate spin-up and spin-down bands, respectively. {\bf a)} $c/a$=1.45 for fixed $c$ and varying $a$, {\bf b)} $c/a$=1.55 and {\bf c)} $c/a$=1.45 for fixed $a$ and varying $c$. } \label{bandstr}
\end{center}
\end{figure}

In this section, we will present \textit{ab initio} calculations to address some behavior observed in the experiments discussed above. Let us start with the discussions on the magnetic anisotropy energy. This is a very delicate property to calculate, since a very dense sampling of the Brillouin-zone is needed in order to get accurate results. We have done careful convergence tests, with respect to the number of k-points, and our final calculation was done for 9216 k-points in the full Brillouin-zone. The final self-consistent result of the MAE is shown in Fig.~\ref{maexch}, as a function of $c/a$-ratio. In Fig.~\ref{maexch} we also show the energy difference between a FM phase and an AFM one, as a function of $c/a$-ratio. Note from the figure that we have made $c/a$ changes while keeping either the $c$-axis or $a$-axis fixed at bulk experimental values ($c$=5.712 \AA~ and $a$=3.71 \AA). The bulk experimental $c/a$-ratio is 1.54, and for this geometry the FM phase is stable (see Fig.~\ref{maexch}) over the AFM phase with $\sim$0.1 eV per formula unit. We also note from the theoretical results that the easy axis lies in the $ab$-plane, and that the $c$-axis is a hard magnetization direction. This result is in good agreement with experimental observations, and the size of the MAE (0.25~meV/f.u.~$\approx$12$\times 10^{5}$ J/m$^{3}$) is rather substantial and well comparable to our experimental results, as can be seen by a comparison to data in Table \ref{tab:anisotropy}, as well as to bulk results.\cite{Blois1963} From Fig.~\ref{maexch}a it is seen that the AFM phase is stable when the $c/a$-ratio is 1.65 for a fixed $c$-axis. A rather interesting feature is displayed in Fig.~\ref{maexch}b in that the dependence on the calculated MAE is rather weak when keeping the $a$-axis fixed and varying the $c$-axis, whereas the dependence is very strong when the $c$-axis is fixed and the $a$-axis is varied. The microscopic reason for the MAE lies in the electronic structure \cite{vanvleck} and in order to analyze the behavior in Fig.~\ref{maexch}, we show in Fig.~\ref{bandstr} the energy bands for different $c/a$-ratio. We also display energy bands when the $c$-axis is varied and the $a$-axis kept constant, and vice versa, for $c/a$-ratios equal to 1.45 and 1.55 (experimental value). It is well known that the contribution to the MAE from the spin-orbit coupling in perturbation theory is from occupied bands just below the Fermi level to unoccupied bands just above the Fermi level. Hence, it is the energy bands around the Fermi level one should inspect if one wants to investigate the MAE changes shown in Fig.~\ref{maexch}. The energy bands in Fig.~\ref{bandstr} demonstrate an overall larger sensitivity when modifying the $a$-axis while keeping the $c$-axis fixed, which is consistent with the data in Fig.~\ref{maexch}. This is particularly pronounced along the symmetry direction $\Gamma \to A \to L \to M$. Here entire segments of the spin-down Fermi surface are removed by the strain, since the spin-down band that for the unstrained case crosses the Fermi level, becomes completely unoccupied. This only happens for a strain where the $c$-axis is fixed.

\begin{figure}[h!]
\begin{center}
\includegraphics[width=0.45\textwidth]{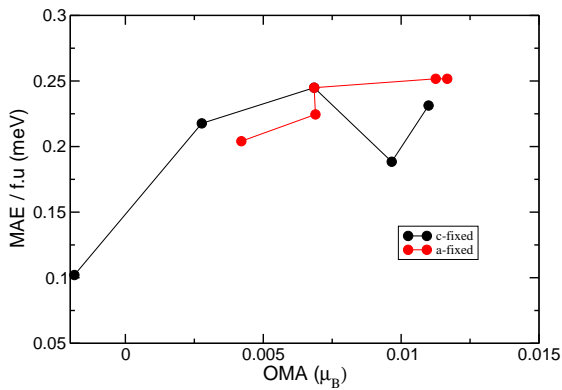}
\caption{(Color online) Magnetic anisotropy energy (MAE) vs. orbital moment anisotropy (OMA) for different $c/a$-ratio. The calculations were done by keeping either the $a$-axis (red circles) or the $c$-axis (black circles) fixed. } \label{maeoma}
\end{center}
\end{figure}

The final theoretical result we would like to discuss is the relationship between MAE and the anisotropy of the orbital moment when measureed along an easy and hard magnetisation direction, the so called orbital moment anisotropy (OMA) shown in
Fig.~\ref{maexch}c. From perturbation theory it has been shown \cite{bruno} that there is a direct coupling between the MAE and the OMA. The derivation is made under the assumption that cross-terms in the spin-orbit interaction (i.e. $\xi (l_+ s_- + l_- s_+)$) can be  neglected compared to diagonal terms (i.e. $\xi l_z s_z$). For a material composed of one element this relationship holds well, and there is a direct proportionality between the MAE and OMA. However, a generalization of this theory to materials composed of more than one element, shows a more complex behavior, \cite{cecilia} since both the MAE and OMA may have contributions from spin-orbit effects of all atomic species of the material. In short, this means that for a material like MnAs, the MAE of the magnetic moment, which is primarily located on the Mn atom, is also influenced by the spin-orbit interaction of the As atom. Since there are cross terms in atomic site index to both the MAE and OMA, which are large when there is strong hybridization of electron states centered on different atomic sites, there is no reason to expect a linear relationship between the MAE and OMA. Figure~\ref{maeoma} shows a comparison between calculated values of the MAE and OMA, for different values of the $c/a$-ratio. Again, the calculations were done by keeping either the $a$-axis or the $c$-axis fixed. Note from the figure that the linear relationship between OMA and MAE is lost. Hence, for this material it is misleading to measure the OMA and from this value try to draw conclusions about the MAE. This has previously been shown experimentally for Au/Co thin films.\cite{cecilia}

\begin{figure}
\begin{center}
\includegraphics[width=0.45\textwidth]{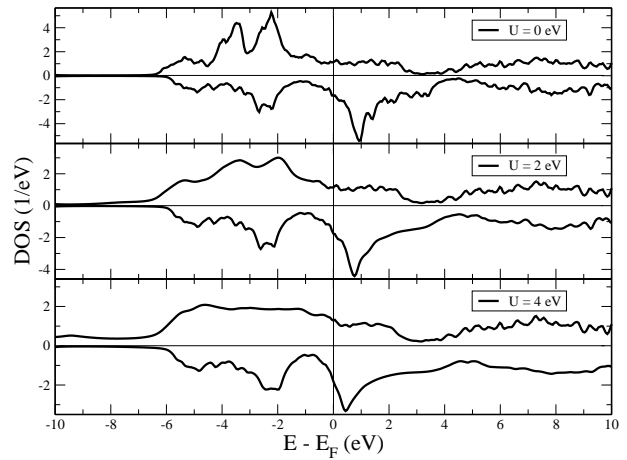}
\caption{Spin-polarized DOS for (top) GGA with U=0 eV (top), (middle) DMFT with U=2 eV and (bottom) DMFT with U=4 eV calculations.} \label{mnasdos}
\end{center}
\end{figure}

The electronic structure of MnAs has been analyzed in the past \cite{ravindran,katoh,oppeneer,sanyal} using LSDA and/or GGA. A
relevant question for this material concerns the influence of electron correlations, which might be important for the electronic structure and hence also for the magnetic properties. Mn compounds often display electronic structures which have a correlated electronic $d$-shell, that may even become localized due to strong electron-electron interaction. Examples of this are LSMO (i.e La$_{1-x}$Sr$_x$MnO$_3$) \cite{solovyev} and MnO.\cite{oka} Electron correlation has recently been discussed even for one of the elemental forms of Mn, the $\gamma$-phase.\cite{dimarco} Hence, it is important to investigate this aspect in MnAs, and for this reason we have compared the electronic structure and the spin and orbital moments of MnAs using GGA and dynamical mean field theory (DMFT - a recent method to take into account electron correlations). \cite{revdmft} In this calculation we have used the implementation by Grechnev \textit{et. al}.\cite{grech} In Fig.~\ref{mnasdos} we compare first the electronic structure from GGA calculations with DMFT calculations, using a Coulomb U (i.e. the strength of the electron-electron interaction) of 0 for GGA and 2 and 4 eV for DMFT. As seen in Fig.~\ref{mnasdos} one can notice some changes in the electronic structure with the variation of the value of U. The variation in the spin-up DOS is more prominent than the spin-down one. However, the resulting exchange splitting does not differ much as discussed below.  Next we turn our attention to the spin- ($m_s$) and orbital ($m_l$) moments. From the GGA calculation we obtain for the Mn $d$-orbital a $m_s$=3.12 $\mu_B$/atom and a $m_l$=0.026 $\mu_B$/atom. From DMFT calculations for U=2~eV we obtain $m_s$=3.12 $\mu_B$/atom and $m_l$=0.028 $\mu_B$/atom,  while for U=4~eV we get $m_s$=3.11 $\mu_B$/atom and $m_l$=0.017 $\mu_B$/atom. Our calculations thus show a rather weak influence on the spin and orbital moments as a function of increasing values of U, implying that  electron correlations outside of what is included in GGA are not of major importance for MnAs. For the GGA calculation we obtain a total moment of 3.13 $\mu_B$/atom, which should be compared to earlier theoretical results of 3.06~$\mu_B$/atom \cite{oppeneer,katoh} and to the bulk experimental value of 3.45~$\mu_B$/atom.\cite{Ido1985}

\section{Discussion}
The experimental $M_s$ values for the As-cap and de-cap samples are at least 10-15\% larger than what has been reported for thin film \cite{Schippan2000} or bulk MnAs.\cite{Ido1985} Values similar to our experimental values were obtained by Sanvito and Hill \cite{Sanvito2000} from calculations of the effect of cell volume expansion where a $M_s$$\sim$4~$\mu_B$ was obtained due to the reduction of $p-d$ hybridization when then unit cell was stretched. The necessary increase of the cell volume required almost unphysical values of $a$ and $c$, but the important observation is the large change in magnetic moment with lattice distortion. Our theoretical results also show that the magnetic properties depend strongly on lattice strain, especially on variations of $a$, see Fig.~\ref{maexch}c. Since XRD is a global probe, the measured $a$ and $c$ correspond to averages over the sample volume and thus locally the tensile strain could be much larger. In fact, our SAED results, where the size of the selected area is 800 nm, indicate are more strained $a$-axis. However, we refrain from depending solely on the TEM results since thinning during sample preparation could induce additional lattice strain. Still, changes in $M_s$ and $A_0$ with annealing show that even small relaxations of the structure play a significant role.          
\newline Our calculated magnetic anisotropy energy and easy axis direction are in good agreement with the experiments. Moreover, the theoretical results suggest that the change in magnetic properties is more sensitive by straining the in-plane axis than the out-of-plane one, an effect which has been attributed to the change in band structures due to strain. The calculated magnetic moments are observed to vary significantly upon straining the system. Following this observation, one may argue that the experimentally measured large moments may arise from the locally strained regions of the samples. In short, our theoretical and experimental results are in line with each other.\\

\section{Conclusions}
In conclusion, our findings indicate a more complex relationship between strain and magnetic properties in MnAs films than a simple stretch/compression of the unit cell axes. A strong dependence of $K_{1}$, $K_{2}$ and $T_{c}$ on the angle and fraction of out-of-plane orientation has been established. The \textit{ab initio} calculations demonstrate a strong dependence of the magnetic properties on the variation of the in-plane axis. We believe that local structural strains that distort the lattice give rise to the enhanced $M_{s}$ and MCA measured in our samples. The possibility to alter the magnetic properties through controlled film growth with maintained epitaxy gives MnAs strong potential as a material for various spintronic applications.

\section{Acknowledgments}
The authors like to thank the Swedish Research Council, Knut and Alice Wallenberg Foundation (KAW), STINT,  Swedish Foundation for Strategic Research (SSF), G\"oran Gustafssson Foundation, Carl Tryggers Foundation for financial support. O.E. is also grateful to the ERC for support.  We also acknowledge Swedish National Infrastructure for Computing (SNIC) for the allocation of time in high performance supercomputers.  


\end{document}